\documentclass[a4paper]{jpconf}

\usepackage{amsmath, amssymb}
\usepackage{graphicx}
\usepackage{color}

\begin{document}
\title{High baryon densities achievable in the fragmentation regions at RHIC and LHC}

\author{Joseph Kapusta and Ming Li}

\address{School of Physics and Astronomy, University of Minnesota, Minneapolis, MN 55455 USA}

\ead{kapusta@physics.umn.edu, ml@physics.umn.edu}

\begin{abstract}
We use the McLerran-Venugopalan model of the glasma energy-momentum tensor to compute the rapidity loss and excitation energy of the colliding nuclei in the fragmentation regions followed by a space-time picture to obtain their energy and baryon densities.  At the top RHIC energy we find baryon densities up to 3 baryons/fm$^3$, which is 20 times that of atomic nuclei.  Assuming the formation of quark-gluon plasma, we find initial temperatures of 200 to 300 MeV and baryon chemical potentials of order 1 GeV.  Assuming a roughly adiabatic expansion it would imply trajectories in the $T-\mu$ plane which would straddle a possible critical point.  
\end{abstract}

For more than three decades the high energy heavy ion community has been focussed on the central rapidity region following the seminal work of Bjorken \cite{BJ1983}.  The main reasons are (i) the energy density is expected to be higher there, (ii) the matter is nearly baryon-free making it relevant to the type of matter that existed in the early universe, and (iii) detectors in a collider can more easily measure particle production and correlations in a few units of rapidity around the center-of-momentum.  The earlier work of Anishetty, Koehler and McLerran \cite{AKM1980}, which found that nuclei were compressed by a factor of 3.5 and excited to an energy density of about 2 GeV/fm$^3$ when they collide at extreme relativistic energies, was pursued only sporadically over the years \cite{C1984,GC1986,MK2002,FS2003,ML2007}.  Here we briefly report on our recent and ongoing work on this subject.  We focus on the baryonic fireballs which emerge at large rapidity in the center-of-momentum frame.  Our calculations are not relevant or applicable to the lower energy beam scans at RHIC, nor to future experiments at NICA and FAIR \cite{CBM}.

Consider central collisions of equal mass nuclei.  We neglect transverse motion, which should not be important during the fraction of a fm/c time interval of relevance.  Then the collision can be thought of as a sum of independent slab collisions each taking place at a particular value of the transverse coordinate ${\bf r}_\perp$ with the beam along the $z$-axis.  The projectile slab has a 4-momentum per unit area in the center-of-momentum frame denoted by ${\cal P}_{\rm P}^{\mu} = ({\cal E}_{\rm P}, 0, 0, {\cal P}_{\rm P})$.  Each slab loses energy and momentum to the classical color electric and magnetic fields produced in the region between the two receding nuclei, referred to as the glasma.  This loss is determined by 
$d{\cal P}^{\mu}_{\rm P} = -T_{\rm glasma}^{\mu\nu}d\Sigma_{\nu}$ where $d\Sigma_{\nu} = (dz,0,0,-dt)$ is the infinitesimal four-vector perpendicular to the hypersurface spanned by $dt$, $dz$, and unit transverse area.  The energy-momentum of the glasma has the form \cite{CFKL2015,LK2016}
\begin{equation}\label{em_tensor}
T^{\mu\nu}_{\rm glasma}=
\begin{pmatrix}
{\cal A}+{\cal B}\cosh{2\eta} & 0 & 0 & {\cal B}\sinh{2\eta} \\
0 & {\cal A} & 0 & 0 \\
0 & 0 & {\cal A} & 0 \\
{\cal B}\sinh{2\eta} & 0 & 0 & -{\cal A}+{\cal B}\cosh{2\eta} \\
\end{pmatrix} \, .
\end{equation}
The $\mathcal{A}$ and $\mathcal{B}$ are known analytical \cite{LK2016} and numerical (for SU(2)) \cite{Gelis:2013rba} functions of proper time $\tau$, while the dependence on space-time rapidity $\eta$ follows from the fact that $T^{\mu\nu}_{\rm glasma}$ is a second-rank tensor in a boost-invariant setting.  The longitudinal position $z_{\rm P}$ at each 
${\bf r}_\perp$ is a function of time.  The $z_{\rm P}(t)$ is related to the time $t$ via the velocity $v_{\rm P} = dz_{\rm P}/dt =\tanh{y_{\rm P}}$, where $y_{\rm P}$ is the momentum-space rapidity of the slab. The Lorentz invariant effective mass per unit area ${\cal M}_{\rm P}$ is defined via the relations ${\cal E}_{\rm P} = {\cal M}_{\rm P} \cosh{y_{\rm P}}$ and ${\cal P}_{\rm P} = {\cal M}_{\rm P} \sinh{y_{\rm P}}$.  The pair of differential equations describe both the loss of kinetic energy of the projectile nucleus and the internal excitation energy imparted to it during the collision.  Thus ${\cal M}_{\rm P}$ is not constant but increases with time, unlike the case of the string model \cite{MK2002}. 

Initial conditions are needed to solve the equations of motion.  For that we turn to hydrodynamical descriptions of collisions at the top RHIC energy of $\sqrt{s_{NN}} = 200$ GeV.  Reference \cite{SBH2011} assumed that viscous hydrodynamics became applicable at $\tau = 0.6$ fm/c with $\varepsilon (r_{\perp}=0, \tau = 0.6 \; {\rm fm/c}) = 30$ GeV/fm$^3$.  Extrapolating back to $\tau = 0$ using the results in \cite{LK2016} gives $\varepsilon_0 \equiv \varepsilon_0 (r_{\perp}=0,\tau=0) =$ 123, 142, and 158 GeV/fm$^3$ for UV cutoffs of $Q=$ 3, 4 and 5 GeV, respectively. We use a Wood-Saxon nuclear density distribution $\rho_{\rm A}$ and assume as usual that the initial energy density is proportional to the square of the thickness function $\int_{-\infty}^{\infty} dz \rho_{\rm A}(r_{\perp},z)$ 

\begin{figure}[h]
\begin{minipage}{18pc}
\includegraphics[width=18pc]{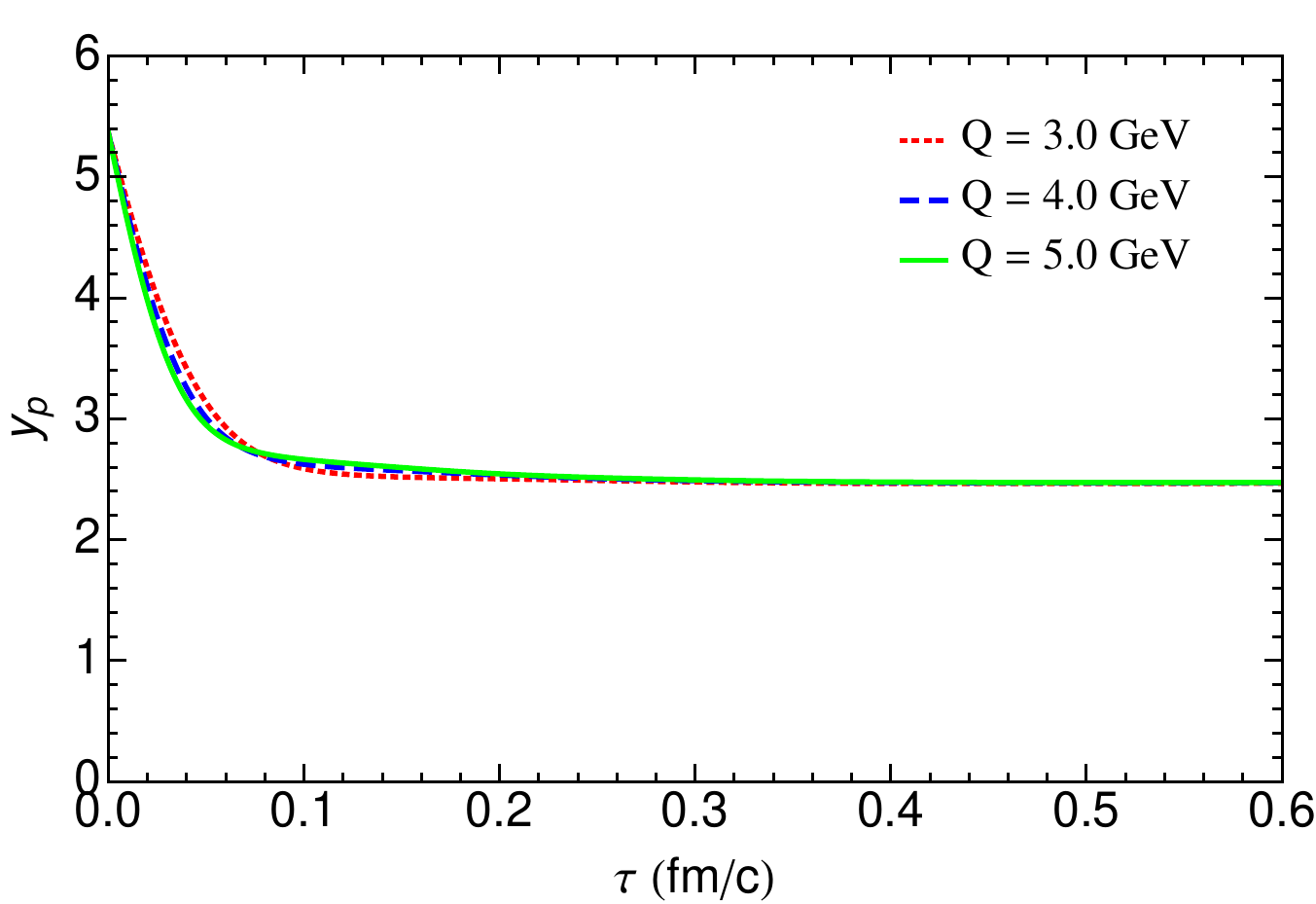}
\caption{\label{F1} Rapidity of the central core of a Au projectile nucleus in the center-of-momentum frame for $\sqrt{s_{NN}} = 200$ GeV as a function of proper time.  The result is insensitive to the choice of $Q$.}
\end{minipage}\hspace{2pc}%
\begin{minipage}{18pc}
\includegraphics[width=18pc]{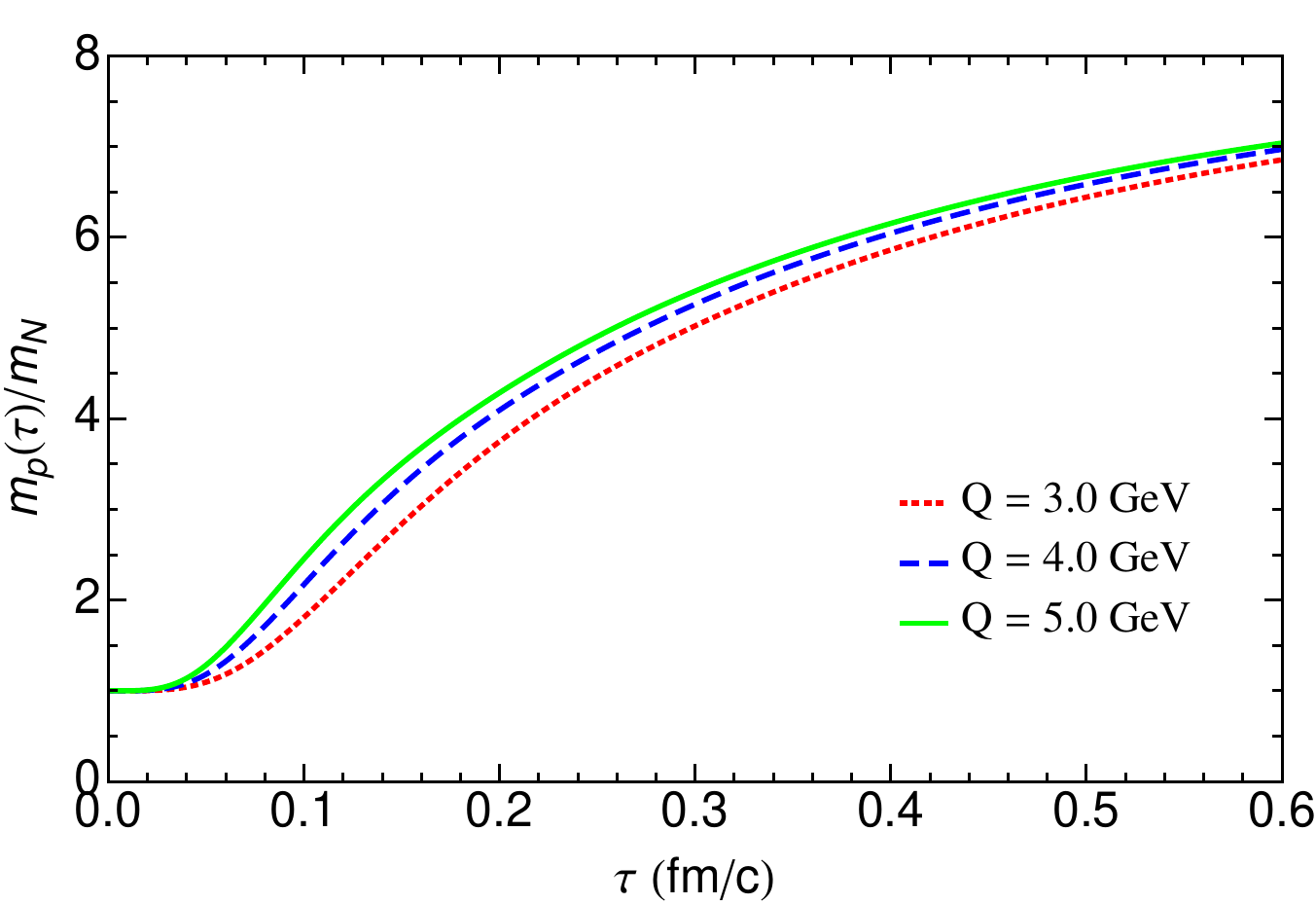}
\caption{\label{F2} Excitation energy per baryon in the central core of a Au projectile nucleus in the center-of-momentum frame for $\sqrt{s_{NN}} = 200$ GeV as a function of proper time.  The result is mildly sensitive to the choice of $Q$.}
\end{minipage} 
\end{figure}

Figure \ref{F1} shows the momentum-space rapidity $y_{\rm P}$ of the central core of a gold nucleus as a function of proper time $\tau$.  The central core loses about 3 units of rapidity within the first 0.1 to 0.2 fm/c; this is a robust result, insensitive to the value of $Q$.  When averaged over the whole nucleus the baryon rapidity loss is about 2.4.  For 0-10\% centrality BRAHMS \cite{BRAHMS2004,BRAHMS2009} found an average rapidity loss of about $2.05 +0.4/-0.6$.  This is consistent with our result, especially since we focus on 0\% centrality for illustration.  Figure \ref{F2} shows the excitation energy per baryon in units of the nucleon mass as a function of proper time.  There is a slow but monotonic increase, unlike the rapidity loss whose asymptotic limit is reached within a few tenths of a fm/c.

\begin{figure}[t]
\begin{minipage}{18pc}
\includegraphics[width=18pc]{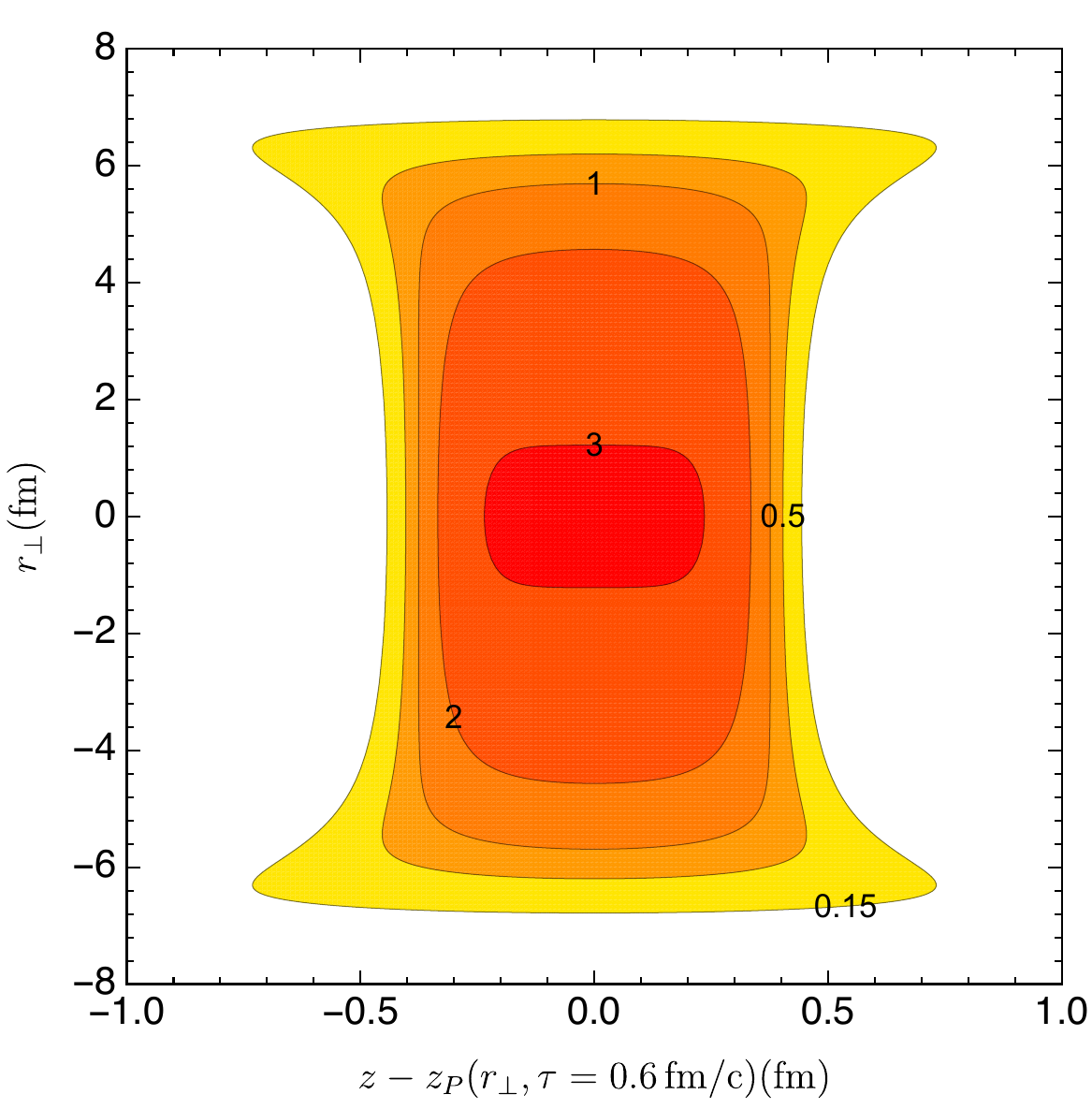}
\caption{\label{F3} Contour plot of the proper baryon density for central collisions of Au nuclei at $\sqrt{s_{NN}} = 200$ GeV.  The units are baryons per fm$^3$.  The horizontal axis measures the distance in the local rest frame.}
\end{minipage}\hspace{2pc}%
\begin{minipage}{18pc}
\includegraphics[width=18pc]{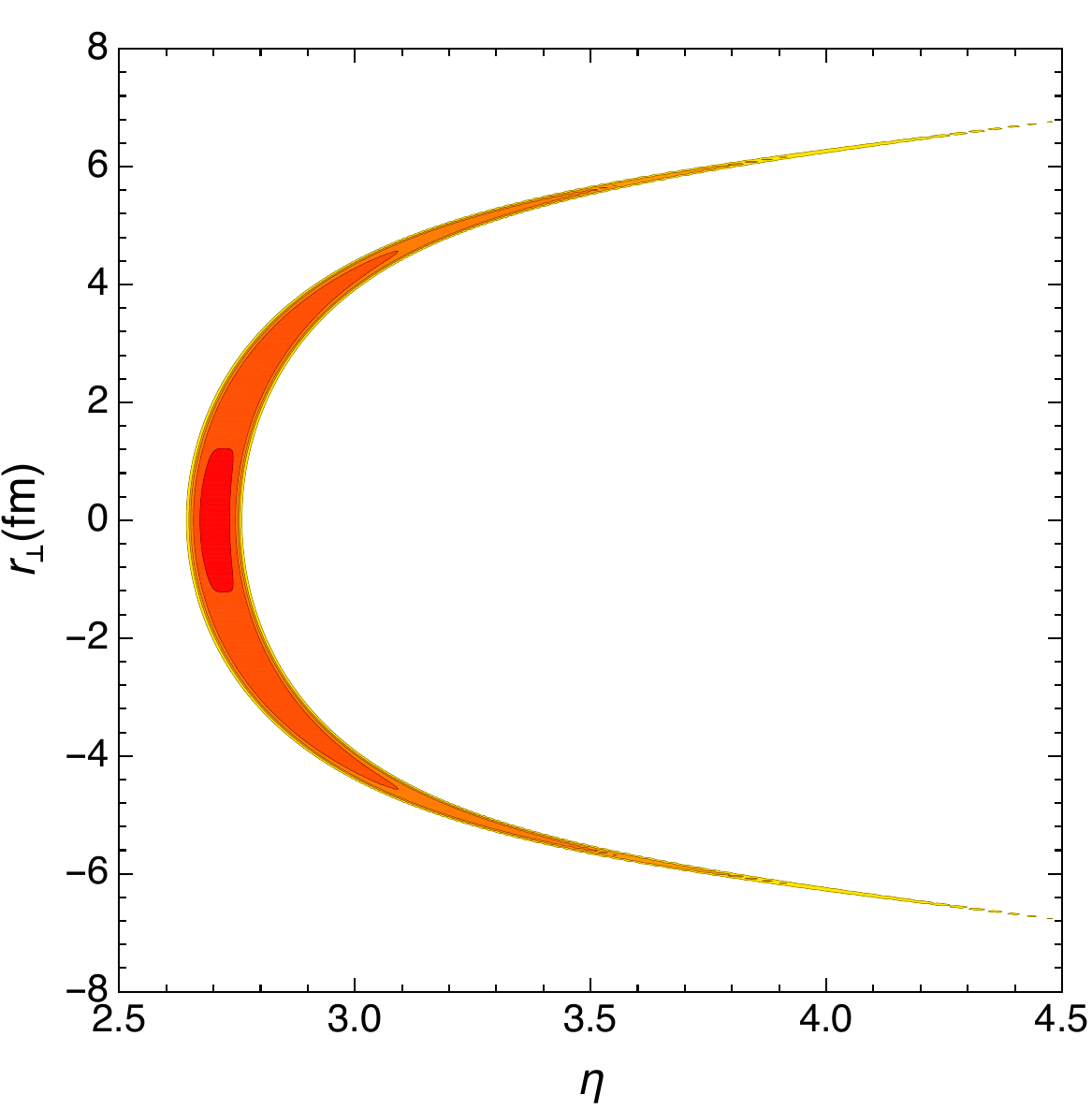}
\caption{\label{F4} Contour plot of the proper baryon density at $\tau = 0.6$ fm/c for central collisions of Au nuclei at $\sqrt{s_{NN}} = 200$ GeV.  The units are baryons per fm$^3$.  The horizontal axis is the space-time rapidity.}
\end{minipage} 
\end{figure}

The McLerran-Venugopalan model assumes that the nuclei can be treated as infinitesimally thin slabs.  It does not address the space-time evolution of the individual nuclei.  Anishetty {\it et al.} \cite{AKM1980} gave a very simple and direct argument that the nuclear matter would be compressed by a factor of $\exp(\Delta y)$ where $\Delta y > 0$ is the rapidity loss (gain) of the projectile (target).  It is a Lorentz invariant quantity which follows from the infinitely thin projectile sweeping through the target in the target rest frame.  The argument was verified in a specific model in \cite{GC1986}.  The local baryon density is
$n_B(r_{\perp},z') = {\rm e}^{\Delta y(r_{\perp})} \rho_{\rm A}\left(r_{\perp}, z'  {\rm e}^{\Delta y(r_{\perp})}\right)$
where $z' = z - z_{\rm P}(r_{\perp})$, all evaluated at $\tau = 0.6$ fm/c.  Contours of the resulting baryon density are shown in fig. \ref{F3} when plotted in the $z$-direction and in fig. \ref{F4} when plotted against space-time rapidity.

The baryon rapidity distribution at $\tau = 0.6$ fm/c is shown in fig. \ref{F5}.  Also shown is the distribution smeared by a Gaussian with a width given by $\sqrt{T/m} = 0.16$.

Assume that the matter in the fireball does equilibrate on the time scale of 0.6 fm/c as argued in {\cite{AKM1980}.  For simplicity consider a massless gas of noninteracting up, down, and strange quarks and gluons.  The net strangeness in the fireball is zero.  Let the up and down quark chemical potentials be equal to each other and to 1/3 of the baryon chemical potential $\mu_{\rm B}$.  The equation of state is then given by the pressure as a function of temperature and baryon chemical potential by
$P(T,\mu_{\rm B}) = \frac{19 \pi^2}{36} T^4 + \frac{1}{9} T^2 \mu_{\rm B}^2 + \frac{1}{162 \pi^2} \mu_{\rm B}^4$. 
The energy density decreases with $r_{\perp}$ faster than the baryon density.  Illustrative numbers are given in Table \ref{T1}.  It would be expected that the hydrodynamic expansion of the fireball would be approximately adiabatic, just as in the central rapidity region.  If that is the case, then the values of the entropy per baryon estimated above would be in just the right range for the trajectories of the fluid elements in the $T-\mu_{\rm B}$ plane to pass near or even through a possible critical point \cite{ABMN2008}.

In conclusion, we have employed the McLerran-Venugopalan model to calculate the energy/rapidity loss of baryons in high energy heavy ion collisions.  We found that the baryon densities in the fireballs outside the central rapidity region attain values an order of magnitude greater than normal nuclear matter.  These findings suggest that further theoretical and experimental studies be done to probe the equation of state at the highest baryon densities achievable in a laboratory setting.

\begin{table}[t]
\caption{\label{T1} Illustrative results for central collisions of Au nuclei at $\sqrt{s_{NN}} = 200$ GeV.  Uncertainties of 33\% in $s$ and $n_B$ result in uncertainties in $T$ and $\mu_B$ of 10\%.}
\begin{center}
\lineup
\begin{tabular}{llllll}
\br
$r_{\perp}$(fm)&$n_B$(baryons/fm$^3$)&$\varepsilon_P$(GeV/fm$^3$)&$T$(MeV)&$\mu_B$(MeV)&$s/n_B$ \\
\mr
\00&3.0&20.0&299&1061&26.2\\
5.25&1.5&\05.5&205&1007&18.9\\
\br
\end{tabular}
\end{center}
\end{table}

\begin{figure}[h]
\begin{minipage}{18pc}
\includegraphics[width=18pc]{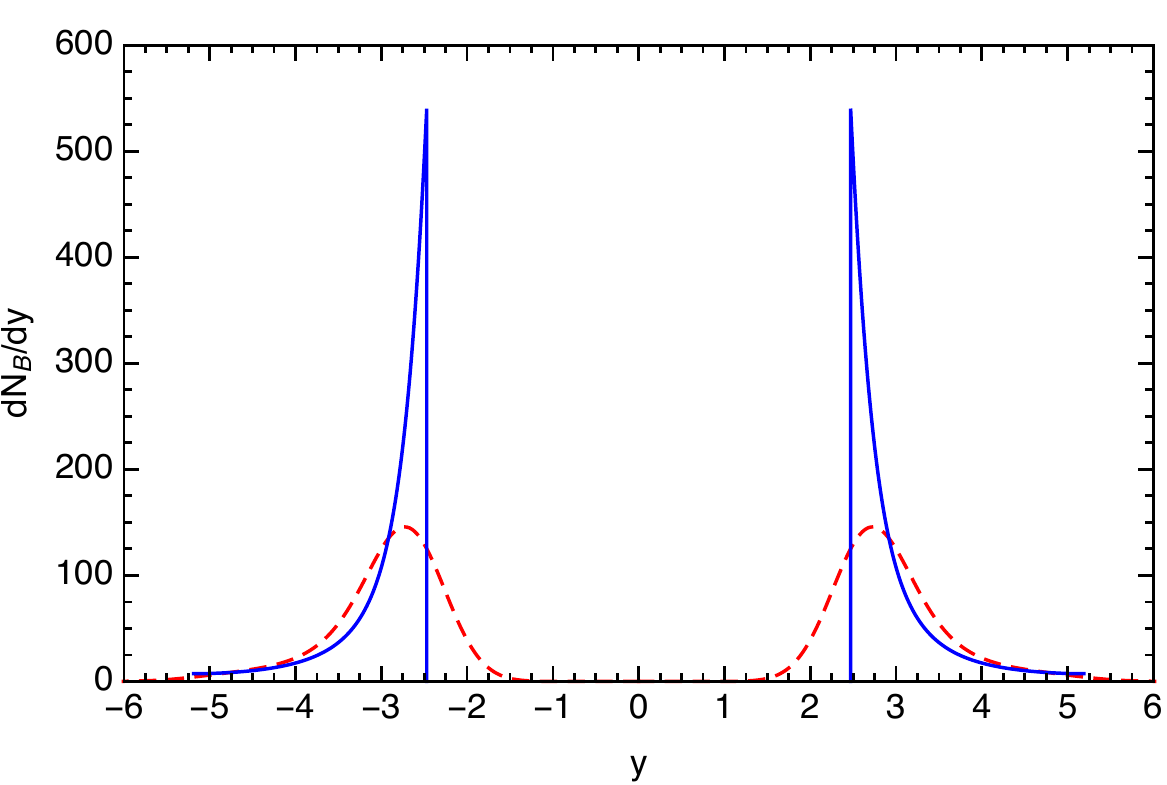}
\caption{\label{F5} Baryon rapidity distribution at $\tau = 0.6$ fm/c with and without thermal smearing.}
\end{minipage}\hspace{2pc}%
\begin{minipage}{18pc}
\includegraphics[width=18pc]{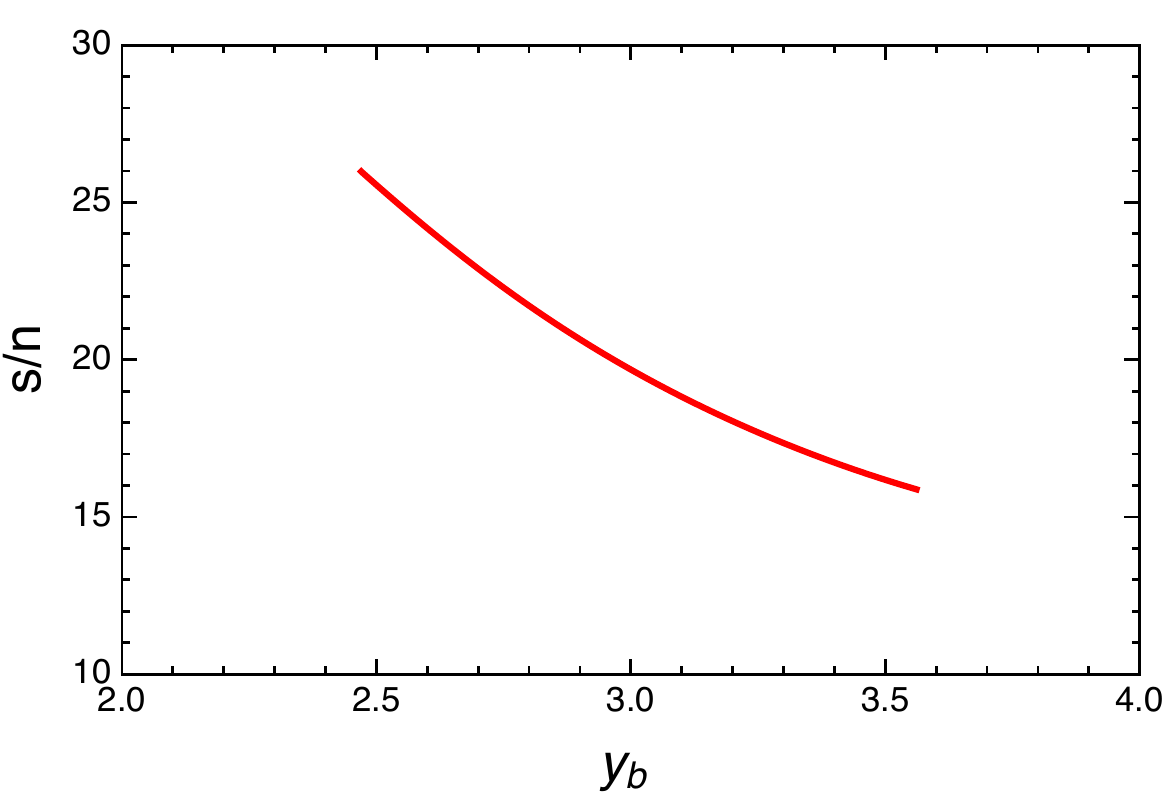}
\caption{\label{F6} Entropy per baryon rapidity distribution at $\tau = 0.6$ fm/c. Beam energy or rapidity scan?}
\end{minipage} 
\end{figure}

\ack This work was supported by the U.S. Department of Energy grant  DE-FG02-87ER40328.


\section*{References}
\begin{thebibliography}{9}

\bibitem{BJ1983}
Bjorken J D 1983 {\it Phys. Rev.} D {\bf 27} 140

\bibitem{AKM1980}
Anishetty R, Koehler P and McLerran L 1980, {\it Phys. Rev.} D {\bf 22} 2793

\bibitem{C1984}
Csernai L P 1984 {\it Phys. Rev.} D {\bf 29} 1945

\bibitem{GC1986}
Gyulassy M and Csernai L P 1986 {\it Nucl. Phys.} A {\bf 460} 723

\bibitem{MK2002}
Mishustin I N and Kapusta J I 2002 {\it Phys. Rev. Lett.} {\bf 88} 112501

\bibitem{FS2003}
Frankfurt L and Strikman M 2003 {\it Phys. Rev. Lett.} {\bf 91} 022301; Strikman M 2006 {\it AIP Conf. Proc.} {\bf 842}, 92

\bibitem{ML2007}
Mishustin I N and Lyakhov K A 2007 {\it Phys. Rev.} C {\bf 76} 011603

\bibitem{CBM}
{\it The CBM Physics Book: Compressed Baryonic Matter in Laboratory Experiments} 2011 ed B Friman {\it et al} (Springer) Lect. Notes Phys. {\bf 814} pp 1-980

\bibitem{CFKL2015}
Chen G, Fries R J, Kapusta J I and Li Y 2015 {\it Phys. Rev.} C {\bf 92}, 064912

\bibitem{LK2016}
Li M and Kapusta J I 2016 {\it Phys. Rev.} C {\bf 94}, 024908

\bibitem{Gelis:2013rba}
Epelbaum T and Gelis F 2013 {\it Phys. Rev. Lett.} {\bf 111}, 232301

\bibitem{SBH2011}
Song H, Bass S A and Heinz U W 2011 {\it Phys. Rev.} C {\bf 83}, 024912

\bibitem{BRAHMS2004}
Bearden I G {\it et al} [BRAHMS Collaboration] 2004 {\it Phys. Rev. Lett.} {\bf 93} 102301

\bibitem{BRAHMS2009}
Arsene I C {\it et al} [BRAHMS Collaboration] 2009 {\it Phys. Lett.} B {\bf 677} 267

\bibitem{coolquarks}
Kurkela A and Vuorinen A 2016 {\it Phys. Rev. Lett.} {\bf 117}, 042501

\bibitem{ABMN2008}
Asakawa M, Bass S A, M\"uller B and Nonaka C 2008 {\it Phys. Rev. Lett.} {\bf 101}, 122302

\end{thebibliography}
\end{document}